\documentclass[12pt]{iopart}
\usepackage{iopams}
\usepackage{setstack}
\usepackage{graphicx}
\usepackage{color}

\newcommand{\bra}[1]{\left< #1 \right|}
\newcommand{\ket}[1]{\left| #1 \right>}
\newcommand{\braket}[2]{\left< #1 | #2 \right>}
\newcommand{\iint}{\int\!\!\!\int}

\begin{document}

\title[Macroscopic superpositions via nested interferometry]{Macroscopic superpositions via nested interferometry: finite temperature and decoherence considerations}

\author{Brian Pepper,$^{1}$ Evan Jeffrey,$^{2}$ Roohollah Ghobadi,$^{3,4}$ Christoph Simon,$^{3}$ and Dirk Bouwmeester$^{1,2}$}
\address{$^{1}$Department of Physics, University of California, Santa Barbara, California 93106, USA\\
$^{2}$Huygens Laboratory, Leiden University, P.O. Box 9504, 2300 RA Leiden, The Netherlands \\
$^{3}$Institute for Quantum Information Science and Department of Physics and Astronomy, University of Calgary, Calgary T2N 1N4, Alberta, Canada \\
$^{4}$Department of Physics, Sharif University of Technology, Tehran, Iran}
\ead{bpepper@physics.ucsb.edu}

\begin{abstract}
Recently there has been much interest in optomechanical devices for the production of macroscopic quantum states. Here we focus on a proposed scheme for achieving macroscopic superpositions via nested interferometry. We consider the effects of finite temperature on the superposition produced. We also investigate in detail the scheme's feasibility for probing various novel decoherence mechanisms.
\end{abstract}


\section{Introduction}
Optomechanical systems have long been investigated as a means of probing the quantum-to-classical transition in macroscopic devices \cite{Bose1999PRA, Marshall2003, Vitali2007,Kleckner2010,Akram2010,Romero2010,Romero2011PRL}. However, it has generally proven difficult to meet all necessary conditions for such experiments. Firstly, a sideband-resolved device is required, allowing ground state cooling \cite{Kleckner2008, Wilson-Rae2007, Marquardt2007, Teufel2011, Chan2011, Kleckner2011}. Secondly, the device's coupling rate must be faster than the mechanical frequency \cite{Marshall2003,Kleckner2008}, in order to create a distinguishable state displaced by more than the device's zero point motion. Finally, the device must meet the strong coupling criterion, ensuring that single photons remain in the cavity long enough to cause significant effects \cite{Marshall2003,Thompson2008,Groblacher2009}. In practice it is very difficult to meet all of these competing requirements simultaneously.

The authors have recently proposed a method to create quantum superpositions in weakly coupled systems via postselected nested interferometry \cite{Pepper2012}. This method greatly relaxes the above requirements, allowing the creation of quantum superpositions with devices more easily in reach of current technology \cite{Kleckner2011}, as well as possible tests of novel decoherence mechanisms \cite{Pepper2012}.

Here we consider the experimental requirements of the proposed nested interferometry scheme, investigating in detail its tolerance of finite temperature in the resonator and finite temperature in the surrounding environment. We also analyze in detail the time scale on which decoherence mechanisms operate, including both traditional environmentally induced decoherence \cite{Zurek2003} and proposed novel decoherence mechanisms \cite{Diosi1989PRA, Penrose1996, Ellis1984, Ellis1989, Ghirardi1986, Ghirardi1990}.
\section{Nested interferometry}
Optomechanical systems evolve under the following Hamiltonian \cite{Law1995}:
\begin{eqnarray}
\hat{\mathcal{H}}=\hbar\omega_o \hat{a}^{\dag}\hat{a} + \hbar\omega_m \hat{c}^{\dag}\hat{c} - \hbar g \hat{a}^{\dag}\hat{a}\left( \hat{c}+\hat{c}^{\dag}\right),\label{eqn:law}
\end{eqnarray}
with $\hbar$ defined as the reduced Planck's constant, $\omega_o$ the optical angular frequency, $\omega_m$ the mechanical angular frequency, the optomechanical coupling rate $g=\omega_o x_{0}/L$, with the zero point motion $x_0=\sqrt{\hbar/(2m\omega_m)}$, $\hat{a}$ the optical annihilation operator, and $\hat{c}$ the mechanical annihilation operator.

If a single photon is input to the cavity and the mechanical state begins in coherent state $\ket{\gamma}_m$, then the mechanical state will evolve as follows \cite{Bose1999PRA}:
\begin{eqnarray}
\ket{\psi(t)}_m &=& \rme^{i\phi(t)} \ket{\gamma(t) + \alpha(t)}_m \\
\gamma(t) &\equiv&\gamma \rme^{-i \omega_m t} \nonumber \\
\alpha(t) &\equiv& \kappa (1 - \rme^{-i \omega_m t}) \nonumber \\
\phi(t) &\equiv& \kappa^2(\omega_m t - \sin \omega_m t), \nonumber
\end{eqnarray}
with $\kappa=g/\omega_m$. Here we define the set of coherent states $\ket{\gamma}$ as well as the single quantum-added coherent states $\ket{\gamma,1}$ \cite{Agarwal1991}:
\begin{eqnarray}
\ket{\gamma} &\equiv& \rme^{-|\gamma|^{2}/2}\sum_{n=0}^{\infty}\frac{\gamma^n}{\sqrt{n!}}\ket{n}\\
\ket{\gamma, 1} &\equiv& \frac{\hat{c}^{\dag}\ket{\gamma}}{\sqrt{\bra{\gamma}\hat{c}\hat{c}^{\dag}\ket{\gamma}}} = \frac{\exp(-|\gamma|^{2}/2)}{\sqrt{|\gamma|^2+1}}\sum_{n=1}^{\infty}\frac{\gamma^{n-1}\sqrt{n}}{\sqrt{(n-1)!}}\ket{n}.
\end{eqnarray}

\begin{figure}[htbp]
\begin{center}
\includegraphics[width=8cm]{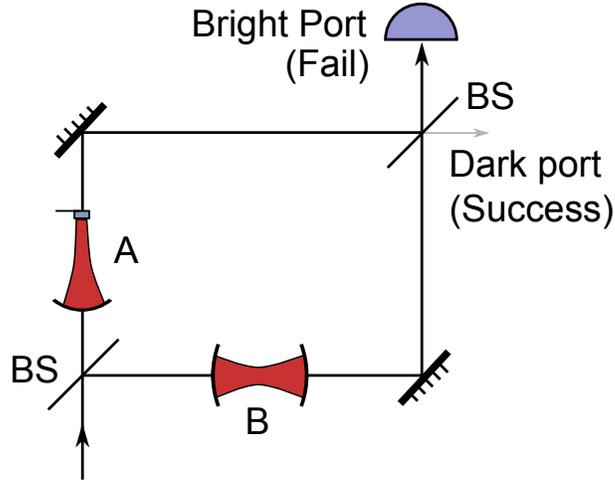}
\caption{The inner interferometer is a Mach-Zehnder interferometer. The upper path contains Cavity A which has a weak optomechanical coupling to a resonator. In the absence of optomechanical interaction the interferometer is balanced and all light exits via the bright port. Postselecting only the photons which exit the normally dark port prepares the resonator in its excited state. Cavity B is used to match the spectrum and time delay of cavity A, and has no optomechanical interaction.}
\label{fig:postsel}
\end{center}
\end{figure}

As detailed in \cite{Pepper2012} the postselection is accomplished by means of an inner Mach-Zehnder interferometer (Fig.~\ref{fig:postsel}). The single photon is input and split into both cavities by a beam splitter, creating state $1/\sqrt{2}(\ket{1}_a\ket{0}_b + \ket{0}_a\ket{1}_b)$. After weakly interacting ($\kappa\ll1$, $\alpha(t) \ll 1$) with the optomechanical resonator for time t, the state will be:
\begin{eqnarray}
\ket{\psi(t)}&=&\frac{1}{\sqrt{2}}\left[\rme^{i\phi(t)} \ket{1}_a\ket{0}_b\ket{\gamma(t) + \alpha(t)}_m + \ket{0}_a\ket{1}_b \ket{\gamma(t)}_m\right].
\end{eqnarray}

By postselecting for photons which exit the dark port, we select the $1/\sqrt{2}(\ket{1}_a\ket{0}_b - \ket{0}_a\ket{1}_b)$ component, and compute it to lowest order in $\kappa$:
\begin{eqnarray}
\ket{\psi_{\mathrm{ps}}(t)}_{m}&=&\frac{1}{2}\left[\rme^{i\phi(t)} \ket{\gamma(t) + \alpha(t)}_m - \ket{\gamma(t)}_m\right] \nonumber \\*
&\approx& \frac{1}{2} \left[ \rme^{-i \kappa \gamma \sin(\omega_m t)} \hat{D}(\alpha(t))-1\right]\ket{\gamma(t)}_m \nonumber \\*
&\approx& \frac{1}{2} \left[(1- i \kappa \gamma \sin(\omega_m t)) (1+\alpha(t)\hat{c}^\dag -\alpha^{*}(t) \hat{c}) -1\right]\ket{\gamma(t)}_m \nonumber \\*
&\approx& \frac{1}{2} \left[\kappa \gamma (1-\cos(\omega_m t))\ket{\gamma(t)}_m + \alpha(t) \sqrt{|\gamma|^2+1}\ket{\gamma(t),1}_m\right] \label{eqn:mechstate}
\end{eqnarray}
with $\hat{D}(\eta)$ defined as the displacement operator. 

In the $\gamma=0$ case, where the resonator has been cooled to its ground state, the above simplifies to a postselected state of:
\begin{eqnarray}
\ket{\psi_{\mathrm{ps}}(t)}_{m}=\frac{\alpha(t)}{2}\ket{1}_{m}.
\end{eqnarray}
Thus in this case, the resonator is placed into the first excited state with probability $|\alpha(t)|^{2}/4$. The weak interaction between the photon and the device is probabilistically amplified.

\subsection{Finite device temperature}
However, for a device of finite temperature, $\gamma\neq 0$. Consider a mechanical resonator initially in a thermal state, a statistical mixture of coherent states:
\begin{eqnarray}
\hat{\rho}_{\mathrm{th}}=\frac{1}{\pi\bar{n}_{\mathrm{th}}}\int \rme^{-|\gamma|^2/\bar{n}_{\mathrm{th}}} (\ket{\gamma}\bra{\gamma}) \rmd^2\gamma \label{eqn:therm}
\end{eqnarray}
where $\bar{n}_{\mathrm{th}}$ is the average number of phonons:
\begin{eqnarray}
\bar{n}_{\mathrm{th}}\equiv\frac{1}{\rme^{\hbar \omega_{m}/ k_{\mathrm{B}} T}-1},
\end{eqnarray}
and where $k_{\mathrm{B}}$ represents the Boltzmann constant. Note that $\bar{n}_{\mathrm{th}}$ is also the value of $|\gamma|^2$ averaged over the thermal distribution, Eqn.~\ref{eqn:therm}. Note that in this subsection we will deal only with mechanical states and will thus drop the $m$ subscript.

For an initial coherent state, we will have created $\ket{\psi_{\mathrm{ps}}(t)}$ from Eqn.~\ref{eqn:mechstate}, a superposition between a small early component with mechanical state $\ket{\psi_{\mathrm{ps}}(t)}$ and a large late component still in $\ket{\gamma(t)}$.

To lowest order in $\kappa$, the overall probability of successful postselection for an initial coherent state will be:
\begin{eqnarray}
\braket{\psi_{\mathrm{ps}}(t)}{\psi_{\mathrm{ps}}(t)} \approx \frac{1}{2}\left[ \kappa^2 (1-\cos \omega_m t) + \frac{1}{2}\kappa^2 |\gamma|^2 \sin^2 \omega_m t \right]\label{eqn:psprob}
\end{eqnarray}

Note that $|\braket{\gamma(t)}{\psi_{\mathrm{ps}}(t)}|^2\approx(1/4)\kappa^2 |\gamma|^2 \sin^2 \omega_m t$, precisely the second term of Eqn.~\ref{eqn:psprob}. Thus the first term represents our signal, while the second term represents a background noise of dark port events due to finite temperature rather than successfully conveying a phonon to the device. Averaging Eqn.~\ref{eqn:psprob} over the thermal distribution, Eqn.~\ref{eqn:therm}, we arrive at:
\begin{eqnarray}
\left<\braket{\psi_{\mathrm{ps}}(t)}{\psi_{\mathrm{ps}}(t)} \right>_{\mathrm{th}}\approx \left[ \kappa^2 \sin^2 \frac{\omega_m t}{2} + \frac{1}{4}\kappa^2 \bar{n}_{\mathrm{th}} \sin^2 \omega_m t \right]\label{eqn:psprobtherm}
\end{eqnarray}
So for the signal to be larger than the noise, we must have $\bar{n}_{\mathrm{th}} \ll 4\left[ \sin(\omega_m t /2)/ \sin \omega_m t\right]^2= \sec^2(\omega_m t / 2)$. This implies that the nested interferometry proposal will only be successful if  $\bar{n} \ll 1$, that is $T\ll \hbar \omega_m / k_B$. Thus, ground state cooling is essential for the success of this scheme. For a sideband-resolved device, this can be accomplished by driving the red (anti-Stokes) sideband of the cavity with a coherent beam \cite{Wilson-Rae2007,Marquardt2007,Teufel2011,Chan2011}. 




\subsection{Nested interferometry}
\begin{figure}[htbp]
\begin{center}
\includegraphics[width=14cm]{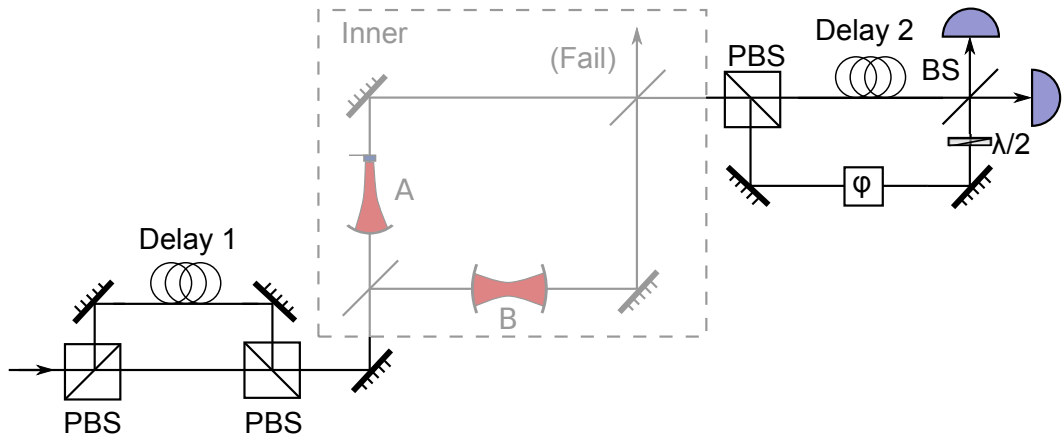}
\caption{The outer interferometer measures the coherence of superposition states by the use of matched time delays. The input pulse is split by polarizing beam splitters (PBS) into an early and late component each of which traverses the inner interferometer (see Fig.~\ref{fig:postsel}). The early and late components are brought back together with a second delay line and the interference visibility is measured by varying the phase shift $\phi$. During the interval between the early and late components the resonator will be in a (postselected) superposition of excited and not-excited, and any decoherence during that time will reduce the final measured visibility.}
\label{fig:readout}
\end{center}
\end{figure}

The nested interferometry proposal \cite{Pepper2012} aims to use this amplification to create macroscopic superposition states, doing so by means of the extended optical setup pictured in Fig.~\ref{fig:readout}. The postselection interferometer of Fig.~\ref{fig:postsel} is nested in a larger interferometer with both an early and a late path.

An experiment begins with the optomechanical device being cooled to its ground state by standard optomechanical cooling techniques \cite{Wilson-Rae2007,Marquardt2007}. Single photons are input to the outer interferometer and are split into an early component and a late component. The late component enters the first delay line. The early component immediately enters the inner interferometer where it interacts with the device, and only the $\ket{1}_m$ component is passed through the dark port, entering a second equal length delay line. 

At this point, the late component is associated with mechanical component $\ket{0}_m$ while the early component is associated with $\ket{1}_m$. These components are left to evolve freely for the length of the delay lines, which can, in principle, be arbitrarily long. During this time they may experience decoherence from either traditional environmentally induced decoherence \cite{Zurek2003} or one of many proposed novel decoherence mechanisms \cite{Diosi1989PRA, Penrose1996, Ellis1984, Ellis1989, Ghirardi1986, Ghirardi1990}. 

Finally, the components exit the delay lines and the late component enters the interferometer. As before only the $\ket{1}_m$ component passes out of the dark port and we are left with both components in the $\ket{1}_m$ state, assuming no decoherence has occurred. At this point both components are interfered to check for visibility, allowing us to measure whether decoherence has taken place.

This scheme has two advantages over previous schemes. First, it allows weakly coupled devices to be placed in superpositions by a single photon. Second, in principle, it allows observation of decoherence on an arbitrary time scale, as the delay lines can be varied. Previous schemes \cite{Marshall2003,Kleckner2008,Akram2010} were limited in the time scales by both the mechanical period of oscillation and the cavity lifetime. This would require new devices to measure at different time scales. Though it may be difficult to determine the cause of the decoherence beyond any doubt, it will be possible to vary the temperature and the characteristics of the device, such as mass, frequency, mechanical quality factor, and optical finesse, allowing parameter dependence to be established.

\section{Decoherence}
Here we will review the various decoherence mechanisms to be considered in this paper. The devices to be considered are hypothetical optomechanical trampoline resonators \cite{Kleckner2011, Pepper2012}, optimized for the nested interference scheme (Tab.~\ref{tab:realdev}). 

\begin{table}[htb]
\caption{We include parameters for two trampoline resonators \cite{Kleckner2011} close to being able to implement the scheme and two devices proposed in \cite{Pepper2012} that should allow the scheme to be implemented. The parameters are effective mass of the mechanical mode (ng), mechanical mode frequency (kHz), cavity length (cm), optical finesse of cavity, mechanical quality factor, environmentally induced decoherence temperature (K), $\kappa=g/\omega_m$, and sideband-resolution measure $\omega_m / \Gamma_c$. Proposed device no. 2 may be capable of observing novel decoherence mechanisms \cite{Ellis1984,Ellis1992,Penrose1996,Diosi1989PRA}.}
\label{tab:realdev}
\begin{tabular}{l r@{}l r@{}l r@{}l r@{}l r@{}l c r@{}l r@{}l}
\br
\multicolumn{1}{c}{Device} & \multicolumn{2}{c}{$m$} & \multicolumn{2}{c}{$f_m$} & \multicolumn{2}{c}{$L$} & \multicolumn{2}{c}{$F$} & \multicolumn{2}{c}{$Q_{m}$} & \multicolumn{1}{c}{$T_{\mathrm{EID}}$} & \multicolumn{2}{c}{$\kappa$} & \multicolumn{2}{c}{$\omega_m / \Gamma_c$}\\
\mr
Tramp. \#1 \cite{Kleckner2011} &  60 && 158 &     & 5 &     &    38,000 &&    43,000 && 0.3 & 0 & .000034 & 2 & .0 \\
Tramp. \#2 \cite{Kleckner2011} & 110 &&   9 & .71 & 5 &     &    29,000 &&   940,000 && 0.4 & 0 & .0016   & 0 & .09 \\
Proposed \#1 \cite{Pepper2012}            &   1 && 300 &     & 0 & .5  &   300,000 &&    20,000 && 0.3 & 0 & .001    & 3 & .0 \\
Proposed \#2 \cite{Pepper2012}            &  100 &&   4 & .5 & 5 &     & 2,000,000 && 2,000,000 && 0.4 & 0 & .005    & 3 & .0 \\
\br
\end{tabular}
\end{table}

\subsection{Environmentally induced decoherence}
Most devices proposed for ground state cooling \cite{Teufel2011,Chan2011,Kleckner2011} require that the device be optically cooled below the temperature $T_{\mathrm{env}}$ that the surrounding environment can reach by conventional cooling (there is one notable exception \cite{OConnell2010}). This is also true of the devices proposed in Tab.~\ref{tab:realdev}.

In this situation, the mechanical resonator is modeled as coupled to an infinite bath of harmonic oscillators \cite{Zurek2003, Kleckner2008}. In the limit of $k_{\mathrm{B}}T_{\mathrm{env}}\gg \hbar \omega_{m}$, mechanical quality factor $Q_m \gg 1$, and a Markovian regime with no memory effects in the bath, the bath degrees of freedom can be eliminated and the system can be described by the master equation for the reduced density matrix $\hat{\rho}$ \cite{Zurek2003, Kleckner2008, Caldeira1983}:
\begin{eqnarray}
\frac{\rmd}{\rmd t}\hat{\rho}= \frac{i}{\hbar}\left[\hat{\rho}, \hat{\mathcal{H}}_{\mathrm{ren}} \right]-\frac{i \gamma_m}{\hbar}\left[\hat{x} , \left\{\hat{p} , \hat{\rho}\right\} \right]-\frac{D}{\hbar^2}\left[\hat{x} , \left[\hat{x}, \hat{\rho} \right] \right],\label{eqn:zurekmaster}
\end{eqnarray}
with $\hat{\mathcal{H}}_{\mathrm{ren}}$ the Hamiltonian from Eqn.~\ref{eqn:law} renormalized by the interaction of the device and the bath, the damping coefficient $\gamma_m=\omega_m/Q_m$, and the diffusion coefficient $D=2m\gamma_{m} k_{\mathrm{B}} T_{\mathrm{env}}$. The first term represents the unitary evolution of the system under the Hamiltonian from Eqn.~\ref{eqn:law}, while the second term represents the damping and the third term represents the diffusion. In the macroscopic regime the diffusion term proportional to $D/\hbar^2$ dominates Eqn.~\ref{eqn:zurekmaster} \cite{Zurek2003, Kleckner2008}. Thus the resulting time scale for decoherence is:
\begin{eqnarray}
\tau_{\mathrm{EID}}\approx\frac{\hbar^2}{D(\Delta x)^2}=\frac{\hbar Q_m}{2k_{\mathrm{B}} T_{\mathrm{env}}},
\end{eqnarray}
with the superposition size $\Delta x=x_{0}$.  It is helpful at this point to define an environmentally induced decoherence temperature \cite{Kleckner2008}:
\begin{eqnarray}
T_{\mathrm{EID}}=\frac{\hbar \omega_m Q_m}{k_B}.
\end{eqnarray}
We note that the inverse of the decoherence time scale is $\tau_{\mathrm{EID}}^{-1}=2\omega_m (T_{\mathrm{env}}/T_{\mathrm{EID}})$. Thus for the environmentally induced decoherence to act on a time scale slower than the mechanical frequency it is necessary that $T_{\mathrm{env}}\ll T_{\mathrm{EID}}$.

We will consider EID with a base temperature of $T_{\mathrm{env}}=1$~mK, obtainable with a dilution refrigerator. For this case, for the $300$~kHz device, $\tau_{\mathrm{EID}}\approx 150$ $\mu$s. For the $4.5$~kHz device, $\tau_{\mathrm{EID}}\approx 15$ ms.

\subsection{Gravitationally induced decoherence}
Gravitationally induced decoherence, proposed independently by Di{\'o}si \cite{Diosi1989PRA} and Penrose \cite{Penrose1996}, is a type of decoherence caused by an object in superposition's perturbation of spacetime. The time scale for such decoherence is:
\begin{eqnarray}
\tau_{\mathrm{P}}=\hbar/\Delta_{\mathrm{P}}
\end{eqnarray}
with the $\Delta_{\mathrm{P}}$ defined as follows:
\begin{eqnarray}
\Delta_{\mathrm{P}} = 4 \pi G \iint \frac{(\rho_1(\vec{x})-\rho_2(\vec{x}))(\rho_1(\vec{y})-\rho_2(\vec{y}))}{|\vec{x}-\vec{y}|}\rmd^{3}x \rmd^{3}y,\label{eqn:penrose}
\end{eqnarray}
with $\rho_1(\vec{x})$ and $\rho_2(\vec{x})$ the mass distributions of the two superposed states.

As in \cite{Kleckner2008}, we model the system as set of spheres representing nuclei. The Penrose energy for one sphere is given by $\Delta_{\mathrm{P}}^0=4\pi(E_{1,2}^0+E_{2,1}^0-E_{1,1}^0-E_{2,2}^0)$, with $E_{m,n}^0= -G\iint\rho_{m}(\vec{x})\rho_{n}(\vec{x})/|\vec{x}-\vec{y}|\rmd^3 x \rmd^3 y$. The spheres considered are far enough apart and displaced little enough that their most significant interaction is with themselves, and not neighboring spheres. This means that we can merely multiply by the number of spheres, $M/m$, to get the total energy $\Delta_{\mathrm{P}}= (M/m)\Delta_{\mathrm{P}}^{0}=4\pi(E_{1,2}+E_{2,1}-E_{1,1}-E_{2,2})$, with $E_{m,n}= (M/m) E_{m,n}^0$

For all cases, we will consider two spherical mass distributions with radii $a$ equal to the size of the specific mass distribution that will be chosen, separated by $\Delta x=x_{0}=\sqrt{\hbar/(2m\omega)}$, the zero point motion of the resonator. Note that this is mathematically equivalent to the model of one sphere at $x=0$ for $\ket{0}_m$, and two half-mass spheres at $x=\pm x_{0}$ for $\ket{1}_m$.

As the radius of the two spheres will be greater than $x_{0}$ regardless of mass distribution used, there will always be significant overlap in the distributions. This will greatly complicate evaluation of Eqn.~\ref{eqn:penrose}. This has no effect on the self-energy terms but does affect the interaction terms. The $1/r$ potential between overlapping spheres has been evaluated previously \cite{Kermode1990}:
\begin{eqnarray}
\fl E_{1,2}=\cases{-GMm/\Delta x & if $\Delta x > 2a$, \\
-GMm\left[ \frac{12a^2-5\Delta x^2}{10a^3} - \frac{\Delta x^5-30\Delta x^3 a^2}{160a^6}\right] & if $0\le\Delta x\le 2a$.}
\label{eqn:energy}
\end{eqnarray}

For the $E_{1,1}$ and $E_{2,2}$ terms, we can just plug $\Delta x = 0$ into Eqn. \ref{eqn:energy}. This gives $E_{1,1}=E_{2,2}=-\frac{6GMm}{5a}$.

There is considerable theoretical disagreement about the proper mass distribution to use for gravitationally induced decoherence \cite{Marshall2003,Diosi2007,Kleckner2008,Maimone2011,Romero2011}. Previous papers have used the zero point motion of the resonator itself, the nuclear radius of the nuclei making up the resonator, the zero point motion of the nuclei making up the resonator, and a completely homogeneous mass with no nuclear granularity. At this point, we will define the mass distributions to be considered in this paper.

\subsubsection{Zero point motion of resonator}
Zero point motion is defined as:
\begin{eqnarray}
a=x_{0}=\sqrt{\frac{\hbar}{2m\omega}}
\end{eqnarray}
For the $300$~kHz device, $a=5.3$~fm. For the $4.5$~kHz device, $a=4.3$~fm.

For this case, for the $300$~kHz device, $\tau_{\mathrm{P}}\approx 3.5$ ms. For the $4.5$~kHz device, $\tau_{\mathrm{P}}\approx 28$ $\mu$s. This type of decoherence might potentially be testable in the $4.5$~kHz device, as it is faster than EID.

\subsubsection{Radius of tantalum}
The atomic nucleus has a size of approximately \cite{Krane}:
\begin{eqnarray}
a=r_0 A^{1/3},
\end{eqnarray}
with $r_0=1.25$~fm and A the atomic mass number. Since the largest component of the mass of a Ta$_2$O$_5$/SiO$_2$ dielectric mirror will be tantalum, we will make the simplifying assumption that the mirrors are composed of tantalum. For tantalum, $A=181$, so $a\approx7$~fm.

For this case, for the $300$~kHz device, $\tau_{\mathrm{P}}\approx 7.1$ ms. For the $4.5$~kHz device, $\tau_{\mathrm{P}}\approx 100$ $\mu$s. This type of decoherence might potentially be testable in the $4.5$~kHz device, as it is faster than EID.

\subsubsection{Zero point motion of nuclei}
In the Debye model, the zero point motion of nuclei in a lattice is given (Eqn. 12.3.10 in \cite{Solyom}):
\begin{eqnarray}
a=x_{0\mathrm{,nuc}}=\frac{3\hbar}{2\sqrt{k_B \Theta_D M}}.
\end{eqnarray}
with $\Theta_{D}$ the Debye temperature and $M$ the atomic mass. Since the largest component of the mass of a Ta$_2$O$_5$/SiO$_2$ dielectric mirror will be tantalum, we will make the simplifying assumption that the mirrors are composed of tantalum. The Debye temperature of tantalum is $\Theta_D=240$~K \cite{Kittel}, and the atomic mass $M=181$~amu. Thus $a\approx5$~pm.

For this case, for the $300$~kHz device, $\tau_{\mathrm{P}}\approx 1.8\times 10^6$ s. For the $4.5$~kHz device, $\tau_{\mathrm{P}}\approx 28\times10^3$ s. This type of decoherence would not be testable, as it is slower than EID in both devices.

\subsubsection{Homogeneous mass}
Some have even proposed modeling the resonator as a perfectly homogeneous mass with no nuclear granularity \cite{Maimone2011, Romero2011}. In general this sets an extremely high bar for the decoherence times, but we will compute it for completeness. In this case we will model the mass as a single sphere of radius $a=30$ $\mu$m (compared to a $60$ $\mu$m diameter cylinder) with mass $60$ ng. It is as though the mirror is composed of one very large nucleus. Though the shape is not correct, this model will suffice for an order of magnitude estimate. This can be represented by setting the nuclear mass $m$ equal to the resonator mass $M$ in Eqn.~\ref{eqn:energy}.

For this case, for the $300$~kHz device, $\tau_{\mathrm{P}}\approx 12\times 10^9$ s. For the $4.5$~kHz device, $\tau_{\mathrm{P}}\approx 1.8\times10^{12}$ s. This type of decoherence would not be testable, as it is slower than EID in both devices.

\subsection{Continuous Spontaneous Localization}
Continuous spontaneous localization is a proposed position-localized decoherence mechanism in which a nonlinear stochastic classical field interacts with objects causing collapse of macroscopic superpositions. Proposed by Ghirardi, Rimini, Weber and Pearle \cite{Ghirardi1986, Ghirardi1990}, the master equation and decay rate for position-localized decoherence have the following form \cite{Romero2011, Penrose1996, Ghirardi1986, Ghirardi1990, Diosi1989PRA, Ellis1989, Ellis1992, Ellis1984}:
\begin{eqnarray}
\frac{\rmd}{\rmd t}\bra{x}\hat{\rho}\ket{x'} &=&\frac{i}{\hbar}\bra{x}[\hat{\rho},\hat{\mathcal{H}}]\ket{x'}-\Gamma(x-x') \bra{x} \hat{\rho} \ket{x'}\label{eqn:master}\\
\Gamma(x)&\equiv&\gamma \left[1- \exp \left(-\frac{x^2}{4a^2}\right)\right] \\
&\approx&\cases{\Lambda x^2 & if $x \ll 2a$, \\
\gamma & if $x \gg 2a$.},
\end{eqnarray}
with $\Gamma(x)$ the decay rate, $\Lambda=\gamma/(4a^2)$ the localization parameter, $\gamma$ the localization strength, and $a$ the localization distance. In all cases, the trampoline resonators considered are in the $x \ll 2a$ limit. For the single nucleon case, the continuous spontaneous localization model \cite{Ghirardi1990} gives values $a_{\mathrm{CSL}}=100$~nm and $\gamma_{\mathrm{CSL}}^0=10^{-16}$~Hz based on phenomenological arguments.

Following \cite{Collett2003, Romero2011}, the value of the localization parameter $\Lambda_{\mathrm{CSL}}$ can be shown to be:
\begin{eqnarray}
\Lambda_{\mathrm{CSL}}=\frac{M^2}{m_0^2} \frac{\gamma_{\mathrm{CSL}}^0}{4 a_{\mathrm{CSL}}^{2}} f(R,b,a)
\end{eqnarray}
with $M$ the resonator mass, $m_0$ the nucleon mass, $R$ the radius of the sphere and $f(R,b,a)$ a parameter depending on the geometry of the device. Disk geometry was considered in \cite{Collett2003}. For motion perpendicular to the disk face $f$ is evaluated (see \cite{Collett2003}, Sec.~5.2, App.~A, and Eqn.~A.11):
\begin{eqnarray}
\fl f(R,b,a)=4\left(\frac{2a}{R}\right)^4\left( \frac{2a}{b} \right)^2 [1-e^{-b^2/4a^2}]\int_{0}^{R/2a}x \rmd x \int_{0}^{R/2a}x' \rmd x' e^{-(x^2+x'^2)}I_{0}(2xx') \label{eqn:fdisk}
\end{eqnarray}
with $R$ the disk radius, $b$ the disk thickness, $I_{0}(x)$ the $n=0$ modified Bessel function of the first kind, and $a$ the localization distance (for CSL, $a_{\mathrm{CSL}}=100$~nm). In the $(R/2a)^2\gg 1$ and $(b/2a)^2\gg 1$ limits, applicable in this case, $f\approx(2a/R)^2(2a/b)^2$.

Thus, for the $300$~kHz device, using a thickness of $\sim 5$~$\mu$m and a radius of $\sim 4$~$\mu$m (values consistent with the proposed finesse and mass), we obtain a decoherence time of order $\tau_{\mathrm{CSL}}=10^7$~s. For the $4.5$~kHz device, using a thickness of $\sim 5$~$\mu$m and a radius of $\sim 40$~$\mu$m, we obtain a decoherence time of order $\tau_{\mathrm{CSL}}=1.5\times10^5$~s. This type of decoherence would not be testable, as it is slower than EID in both devices.

\subsection{Quantum gravity}
It has been proposed that quantum gravity might cause a form of position-localized decoherence due to coupling of the system to spacetime foam. This was first proposed by Ellis, Nanopoulos, Hagelin, and Srednicki \cite{Ellis1984} and subsequently elaborated \cite{Ellis1989,Ellis1992} with others. Notably, this model is phenomenologically equivalent to the CSL model with altered values for the constants \cite{Romero2011}: $a_{\mathrm{QG}}=\hbar m_{\mathrm{P}}/2c m_0^2$ with $m_{\mathrm{P}}=\sqrt{\hbar c/G}$ the Planck mass, and $\gamma_{\mathrm{QG}}^0=4 a_{\mathrm{QG}}^2 c^4 m_0^6 / \hbar^3 m_{P}^3$. This gives us:
\begin{eqnarray}
\Lambda_{\mathrm{QG}} = \frac{M^2}{m_0^2}\frac{\gamma_{\mathrm{QG}}^0}{4a_{\mathrm{QG}}^2}f(R,b,a)= \frac{c^4 M^2 m_0^4}{\hbar^3 m_{P}^3}f(R,b,a)
\end{eqnarray}
with $f(R,b,a)$ as in Eqn.~\ref{eqn:fdisk}. However, since $R\ll a_{\mathrm{QG}}$ and $b\ll a_{\mathrm{QG}}$, we can set $f$ to 1 \cite{Collett2003}:
\begin{eqnarray}
\Lambda_{\mathrm{QG}} \approx \frac{c^4 M^2 m_0^4}{\hbar^3 m_{P}^3}
\end{eqnarray}

Thus, for the $300$~kHz device, using a thickness of $\sim 5$~$\mu$m and a radius of $\sim 4$~$\mu$m, we get a decoherence time of order $\tau_{\mathrm{QG}}=7.1$~s. For the $4.5$~kHz device, using a thickness of $\sim 5$~$\mu$m and a radius of $\sim 40$~$\mu$m, we get a decoherence time of order $\tau_{\mathrm{QG}}=1.1$~ms.  This type of decoherence might potentially be testable in the $4.5$~kHz device, as it is faster than EID.

\section{Conclusion}
In conclusion, we have presented an analysis of the experimental requirements of the nested interferometry scheme \cite{Pepper2012}. The scheme allows for the creation of macroscopic superpositions in weakly coupled systems, and allows for investigation of their decoherence on arbitrary time scales limited only by external delay lines. In particular, we investigate the temperature dependence of the scheme and find that ground state cooling is necessary for implementation. We also investigate the time scales on which proposed novel decoherence mechanisms would be expected to operate. We conclude that two proposed versions of gravitationally induced decoherence \cite{Diosi1989PRA, Penrose1996} are testable, and that quantum gravitational decoherence \cite{Ellis1984, Ellis1989} is testable by this scheme. 

\section*{Acknowledgments}
The authors gratefully acknowledge support by the National Science Foundation grant PHY-0804177, Marie-Curie EXT-CT-2006-042580, European Commission Project MINOS, NWO VICI grant 680-47-604, an AITF New Faculty Award, and an NSERC Discovery Grant.
\section*{References}
\bibliographystyle{unsrt}
\bibliography{njpnested}

\end{document}